\documentclass[reprint,superscriptaddress]{revtex4-2}

\usepackage{amssymb,amsmath}
\usepackage{graphicx}
\usepackage{multirow}
\usepackage{appendix}
\usepackage{mathtools}
\usepackage{amsmath,bm}
\usepackage{siunitx}
\usepackage{lineno}
\usepackage{physics}
\usepackage{wrapfig}
\modulolinenumbers[5]
\graphicspath{{./figures/}}

\begin{document}

\title [mode = title]{Micro-environment of the Eu interstitial in $\beta$-SiAlON:Eu$^{2+}$ green phosphor}

\author{Julien Bouquiaux}
\email{julien.bouquiaux@matgenix.com}
\affiliation{Institute of Condensed Matter and Nanosciences, Universit\'{e} catholique de Louvain, Chemin des \'{e}toiles 8, bte L07.03.01, B-1348 Louvain-la-Neuve, Belgium}
\affiliation{Matgenix, A6K Advanced Engineering Centre, B-6000 Charleroi, Belgium}
\author{Samuel Ponc\'{e}}
\affiliation{Institute of Condensed Matter and Nanosciences, Universit\'{e} catholique de Louvain, Chemin des \'{e}toiles 8, bte L07.03.01, B-1348 Louvain-la-Neuve, Belgium}
\affiliation{WEL Research Institute, avenue Pasteur 6, 1300 Wavre, Belgium}
\author{Yongchao Jia}
\affiliation{Yanshan University,Hebei Key Laboratory of Applied Chemistry, Yanshan University, Hebei Street 438, 066004, Qinhuangdao, P. R. China}
\author{Masayoshi Mikami}
\affiliation{Analysis Technology Laboratory, Science $\&$ Innovation Center, Mitsubishi Chemical Corporation, 1000,
Kamoshida-cho Aoba-ku, Yokohama, 227-8502, Japan}
\author{Xavier Gonze}
\affiliation{Institute of Condensed Matter and Nanosciences, Universit\'{e} catholique de Louvain, Chemin des \'{e}toiles 8, bte L07.03.01, B-1348 Louvain-la-Neuve, Belgium}

\date{\today}

\begin{abstract}
The precise atomic-scale structure around Eu$^{2+}$ activators in the $\beta$-Si$_{6-z}$Al$_z$O$_z$N$_{8-z}$:Eu$^{2+}$ commercial green phosphor remains elusive. 
We use the first-principles $\Delta$SCF excited-state method, embedding of the interatomic force constants for supercells up to 3501 atoms, and Huang-Rhys theory to clarify this issue. 
Monte Carlo exploration is used to identify representative low-energy structural models spanning different levels of Al/O concentration $z$. 
For the lowest-energy structure at low $z$, our computed photoluminescence spectrum reproduces the experimental vibronic peaks at 6~K with excellent agreement in peak positions and intensities, validating the Eu-N$_9$ coordination model with Al, O, and Eu confined to the same crystallographic plane.
Analysis of the low-energy structures reveals that the electron-phonon coupling is weak ($S \approx 2.15$) with a robust characteristic phonon signature across different Al/O arrangements, explaining the surprising persistence of resolved phonon replicas with increasing $z$. 
We explain the experimentally observed red-shift of emission with increasing $z$ through systematic trends in zero-phonon line energies, modest increases in Huang-Rhys factors, and larger configurational diversity at higher compositions. 
\end{abstract}
\maketitle

%%%%%%%%%%%%%%%%%%%%%%%%%%%%%%%%%%%%%%%%%%%%%%%
\section{Introduction}
\label{sec:Intro}

$\beta$-SiAlON:Eu$^{2+}$ stands out as an important commercial green-emitting phosphor material due to its high quantum efficiency, narrow emission bandwidth, and thermal stability, making it particularly suitable for high-power light-emitting diode applications and liquid crystal display backlighting~\cite{nakamura2013blue, xie2009wide, hirosaki2005characterization, wang2025narrow}.
The $\beta$-SiAlON structure is obtained from $\beta$-Si$_3$N$_4$, with Al substitution for Si and O substitution for N, maintaining the charge neutrality of the host lattice. 
This results in the general formula Si$_{6-z}$Al$_z$O$_z$N$_{8-z}$, where $z$ controls the degree of substitution. 
Doping with Eu$^{2+}$ introduces green luminescence, driven by the 5d-4f transitions. 
The characteristic emission peaks around 535~nm, with a full width at half maximum (FWHM) ranging between 45 and 55~nm, which is essential for high-color-purity applications~\cite{nakamura2013blue, hirosaki2005characterization, xie2007synthesis}. 

Despite its technological significance, a fundamental understanding of the structure-property relationships in this material remains incomplete, limiting further improvements. 
In particular, the precise local coordination environment of Eu$^{2+}$ in $\beta$-SiAlON remains an open question. 
Due to its size, it has been recognized early that the Eu$^{2+}$ ion should occupy an interstitial site~\cite{brgoch2014local}. 
Scanning transmission electron microscopy confirmed that this ion is located within the large hexagonal channels of the $\beta$-Si$_3$N$_4$ framework~\cite{kimoto2009direct}. 
However, experimental information about the detailed arrangement of O and Al atoms around the Eu optical activator in $\beta$-Si$_3$N$_4$ remains difficult to obtain due to the almost indistinguishable scattering factors of the two substitution pairs, Al$^{3+}$/Si$^{4+}$ and O$^{2-}$/N$^{3-}$, and the low concentration of the Eu activator~\cite{brgoch2014local}.

In 2016, Wang \textit{et al.}~\cite{wang2016elucidating} performed first-principles calculations aimed at establishing chemical design rules for $\beta$-SiAlON:Eu$^{2+}$. 
Their results indicated that the energetically preferred Eu$^{2+}$ site within the hexagonal channel corresponds to a Eu–N$_9$ coordination polyhedron. 
In that work, several structural models were constructed for different $z$ values. 
With respect to the Al/O ordering, they demonstrated that the total energy is minimized when the following conditions are satisfied: 
(i) all O atoms occupy the 2c Wyckoff positions, 
(ii) each O atom has at least one Al atom as a nearest neighbor, and 
(iii) all Al, O, and Eu species are confined to the same (001) $ab$-plane. 
These rules were later linked to the red shift in emission with temperature~\cite{cozzan2017understanding}. 

Experimental work confirmed the Eu–N$_9$ coordination hypothesis~\cite{takeda2022structure}. 
However, direct experimental validation of the precise micro-environment around the Eu atom remains absent, motivating alternative structural validation accessible with first-principles calculations.
Several studies have measured the photoluminescence (PL) spectrum for various $z$ concentrations and temperatures~\cite{takahashi2012origin, zhang2017controlling, zhang2025narrow}. 
One key feature of the measured spectra is the presence of well-defined vibronic peaks at low $z$ values. 
They are well resolved at low and room temperatures, a seldom observed feature in Eu-doped phosphors. 
This suggests a weak electron-phonon coupling regime (small Huang-Rhys factor). 
This vibronic fingerprint encodes detailed information about local coordination geometry, phonon modes, and electron-phonon coupling that is directly accessible through first-principles calculations. 
Unlike energetic predictions alone, reproducing the experimental vibronic spectrum, including peak positions and relative intensities, provides unambiguous structural validation. 

In this work, we perform structural validation using first-principles vibronic calculations. 
We generate low-energy structural models of $\beta$-SiAlON:Eu$^{2+}$ based on Monte Carlo exploration of the Al/O configuration space, for different $z$ values. 
We then compute Eu$^{2+}$ 5d$\to$4f PL lineshapes for these structural models with a first-principles approach based on the $\Delta$SCF constrained occupation method~\cite{gunnarsson1976exchange, jones1989density, jia2017first, bouquiaux2021importance} with a multi-phonon mode Huang-Rhys model~\cite{huang1950theory, alkauskas2014, jin2021photoluminescence}. 
Comparison between the spectra obtained with this first-principles technique and experiment allows us to validate the structural models. 

Our exploration of 14 representative low-energy structures (12 ab-planar and 2 out-of-plane) covering three compositions ($z$ = 0.1875, 0.25, 0.3125) reveals several key findings. 
First, the computed PL lineshape for the most stable structure at $z$ = 0.1875 shows excellent agreement with experimental vibronic peaks at 6~K, validating the Eu-N$_9$ coordination model with in-plane Al/O arrangement. 
Second, the weak electron-phonon coupling and characteristic vibronic signature are remarkably robust across different Al/O configurations, explaining the persistence of resolved phonon replicas. 
Third, we establish the microscopic origin of composition-dependent trends including the systematic red-shift of emission with increasing $z$ and the appearance of spectral shoulders from energetically accessible structural variants. 

This work is organized as follows. 
We first present the Monte Carlo approach to generate low-energy structural models and the methodology used to obtain the PL lineshape. 
We then analyze the electronic structure, phonon modes, Huang-Rhys spectral function of the lowest-energy model at low $z$, and compare the results with experimental PL spectrum at 6~K.
Finally, we present an overview of the computed lineshapes across all structures, establish $z$-dependent trends, and discuss the implications of these results in light of the known experimental PL spectra.

\section{Computational methodology}
\label{Sec:method}
\subsection{Structural Model Generation via Monte Carlo Exploration}

To explore the configurational space of Al/O substitutions in $\beta$-Si$_{6-z}$Al$_z$O$_z$N$_{8-z}$:Eu, we use a Monte Carlo-based sampling approach with simulated annealing~\cite{kirkpatrick1983optimization,cordell2021probing,vinograd2007coupled}. 
We use the MACE (\textsc{MACE-MP-0a large}) machine-learned interatomic potential~\cite{batatia2022mace,batatia2025design} for fast energy evaluation.
Starting from a 2$\times$2$\times$4 supercell of hexagonal $\beta$-Si$_3$N$_4$ (P6$_3$/m, 224 atoms) with one interstitial Eu atom, we perform stochastic sampling through substitutional swap moves (Al$\leftrightarrow$Si, O$\leftrightarrow$N) accepted with a probability that favors energetically favorable configuration while still allowing occasional higher energy moves to ensure proper exploration of the system's configurational space.
In the literature, $\beta$-SiAlON:Eu$^{2+}$ is often written as Si$_{6-z}$Al$_z$O$_z$N$_{8-z}$:Eu$^{2+}$. However, the introduction of the divalent Eu$^{2+}$ activator requires additional charge compensation. 
Assuming that charge neutrality is achieved by adjustments in the O/N anion ratio, we allow the number of Al and O substitutions to differ, which can be expressed as Si$_{6-z}$Al$_z$O$_{z-2y}$N$_{8-z+2y}$:$y$Eu$^{2+}$. 
Consequently, for the 224-atom supercell considered here, the lowest accessible composition contains 1 Eu, 1 O, and 3 Al atoms ($z$ = 0.1875), and higher substitution levels were explored with 1 Eu; 2 O; 4 Al ($z$ = 0.25) and 1 Eu; 3 O; 5 Al ($z$ = 0.3125 ) configurations.

Monte Carlo sampling revealed a strong energetically-driven tendency for Al-O aggregation, with Eu atoms preferentially located near Al-O-rich regions along the hexagonal channel. 
We confirm that structures with Al, O, and Eu species confined to the same (001) $ab$-plane are energetically favoured~\cite{wang2016elucidating}.
For each target composition ($z$ = 0.1875, 0.25, 0.3125), we select the four lowest-energy configurations, generating 12 representative model structures, all presenting this ab-planar configuration. Two additional energetically unfavoured configurations (one with an out-of-plane Al atom and one with the Al$_3$O cluster far from Eu) are also included in this study to further assess the sensitivity of the vibronic properties to the Al/O arrangement.

All selected structures were then relaxed with density functional theory (DFT). 

Detailed information on the Monte Carlo methodology is provided in section S1 of the Supporting Information.
 
\subsection{Photoluminescence Lineshape Calculations}

Within the Huang-Rhys theory and the Franck-Condon approximation~\cite{huang1950theory,lax1952franck}, the luminescence intensity as a function of photon energy $\hbar\omega$ is given by:
\begin{equation}
L(\hbar\omega,T) \propto \omega^3 A(\hbar\omega,T),
\label{eq:LA}
\end{equation}
where the lineshape function $A(\hbar\omega,T)$ is obtained as the Fourier transform of the
generating function $G(t,T)$~\cite{kubo1955application, alkauskas2014, jin2021photoluminescence}:
\begin{align}\label{eq:L(hw)_generating}
A(\hbar\omega, T) &=  \int_{-\infty}^{+\infty} G(t,T)e^{i\omega t-\frac{\gamma}{\hbar}\abs{t}-i\frac{E^{\rm ZPL}}{\hbar}t}dt,\\
G(t,T)            &= e^{S(t)-S(0)+C(t,T)+C(-t,T)-2C(0,T)}. \label{eq:G_generating}
\end{align}
Here, $\gamma$ is an homogeneous Lorentzian broadening,
$E^{\rm ZPL}$ is the zero-phonon line energy, and
$S(t) = \sum_{\nu} S_\nu\, e^{i\omega_{\nu}t}$ and
$C(t,T) = \sum_{\nu} \overline{n}_\nu(T)\,S_\nu\, e^{i\omega_{\nu}t}$
are the Fourier transforms of the Huang-Rhys spectral function and its
temperature-weighted counterpart, respectively.
$\overline{n}_\nu(T)$ is the Bose-Einstein occupation number of
the $\nu$-th phonon mode:
\begin{equation}
    \label{eq:average_phonon}
    \overline{n}{_\nu}(T)=\frac{1}{e^{\frac{\hbar\omega_{\nu}}{k_{\rm B}T}}-1}.
\end{equation}
The Huang-Rhys spectral function,
$S(\hbar\omega) = \sum_{\nu} S_{\nu}\,\delta(\hbar\omega - \hbar\omega_{\nu})$, quantifies the electron-phonon coupling strength resolved over phonon modes.
The partial Huang-Rhys factor $S_\nu$, which measures the coupling strength of the electronic transition to a single phonon mode $\nu$, reads:
\begin{equation}
    S_{\nu} = \frac{\omega_{\nu}\,\Delta Q_{\nu}^2}{2\hbar},
    \label{eq:S_nu}
\end{equation}
where $\Delta Q_\nu$ is the mass-weighted atomic displacement along phonon mode $\nu$. 
Under the harmonic approximation, this can be expressed in terms of forces:
\begin{equation}
\Delta Q_\nu=\frac{1}{\omega_\nu^2}\sum_{\kappa}\frac{\Delta \mathbf{F}_{\kappa} \mathbf{e}_{\nu,\kappa}}{\sqrt{M_{\kappa}}},
\label{eq:Delta_Q_nu_forces}
\end{equation}
where $\Delta \mathbf{F}_{\kappa}$ are the ground-state forces of atoms $\kappa$ evaluated at the equilibrium excited-state structure, $\mathbf{e}_{\nu,\kappa}$ are the phonon eigenvectors, and $M_{\kappa}$ are the atomic masses. 
This force-based formulation is advantageous as forces decay faster than displacements with distance from the defect \cite{bouquiaux2023first,turiansky2025approximate,jin2021photoluminescence}.

\subsection{Density functional theory calculations}
We use DFT with \textsc{ABINIT} software~\cite{gonze2002first, verstraete2025abinit} and the projector augmented wave (PAW) method~\cite{torrent2008implementation}.
The generalized gradient approximation (GGA-PBE)~\cite{perdew1996generalized} treats exchange-correlation effects, with a Hubbard $U$=7~eV term~\cite{amadon2008gamma, anisimov1991density} on the Eu 4f states, consistent with our previous works~\cite{jia2017first, bouquiaux2021importance, bouquiaux2023first}. 
Structures are relaxed below a maximal residual force of $10^{-5}$~Ha/Bohr with a kinetic energy cutoff of 25~Ha and a single zone-centered $\mathbf{k}$-point. 
For the Eu 4f$^6$5d$^1$ excited state, we use the $\Delta$SCF method with constrained occupation of Eu 4f and 5d states~\cite{gunnarsson1976exchange, jones1989density, jia2017first, bouquiaux2021importance}. 
The occupation of the highest Eu $4f$ state is forced to be unoccupied while the next Eu $5d$
energy state is constrained to be occupied. 
The ZPL energy is computed as the energy difference between the relaxed excited states and the ground state.

Phonon modes of the undoped system (hexagonal $\beta$-Si$_3$N$_4$, 14 atoms per unit cell) are obtained with DFPT~\cite{gonze1997dynamical} with a 5$\times$5$\times$10 $\mathbf{q}$-grid.
Phonon modes of the doped system are computed using finite differences with \textsc{Phonopy}~\cite{togo2015first, togo2023first}. 
To reach the dilute limit, we use the IFCs embedding procedure~\cite{alkauskas2014, jin2021photoluminescence, razinkovas2021vibrational, bouquiaux_2026_18756677} for supercells up to 3501 atoms. 
For the reference structure (section \ref{subsec:ref_stru}), the IFCs are computed with the above-mentioned DFT procedure. 
For the other structural models, the IFCs matrix is computed using \textsc{Mattersim} machine-learned interatomic potential~\cite{yang2024mattersim, sharma2025accelerating, loew2025universal, zhou2025one, turiansky2025machine}. 
We use the dedicated \textsc{Lumabi} package~\cite{bouquiaux2026lumabi} to help with the pre and post-processing of all calculations. 
\textsc{Lumabi} is a Python package within the \textsc{AbiPy}~\cite{verstraete2025abinit} framework that automates the computation of phonon-resolved luminescence spectra of point defects and dopants. 
Convergence of $S(\hbar\omega)$ with supercell size and a direct comparison between DFT and \textsc{Mattersim} IFCs for the reference structure are provided in S2 of the Supporting Information.

\section{Results and discussion}

\begin{figure}[t]
	\centering
	\includegraphics[width=0.85\linewidth]{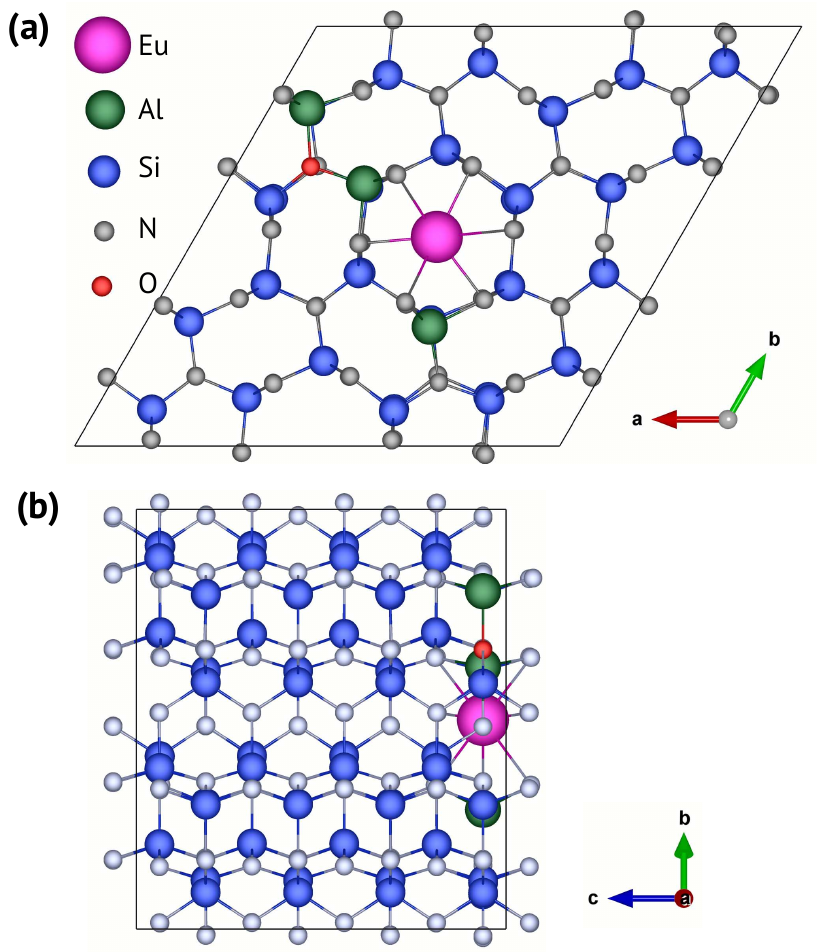}
	\caption{
    Lowest-energy configuration of a 225 atoms supercell with composition Si$_{93}$Al$_3$ON$_{127}$:Eu ($z$=0.1875), selected via Monte Carlo exploration and relaxed with DFT, as described in Sec.~\ref{Sec:method}. 
    The Eu atom occupies an interstitial site within the empty channel running along the $c$ axis, in a distorted Eu-N$_9$ environment. 
    Eu, Al, and O atoms are coplanar in the $ab$ plane, as seen from the $c$ and $a$ axes in panels~(a) and~(b).}
	\label{fig:ref_stru}
\end{figure}

We start with the representative structure shown in Fig.~\ref{fig:ref_stru}. 
This reference structure corresponds to the lowest-energy configuration at composition $z$=0.1875 (one oxygen substitution), selected via Monte Carlo exploration as described in Sec.~\ref{Sec:method}. 
We then extend the analysis to 14 structures covering different $z$ and Al/O configurations to assess the sensitivity of the photoluminescent properties to $z$ and Al/O configurations.

\begin{figure*}[t]
	\centering
	\includegraphics[width=0.98\textwidth]{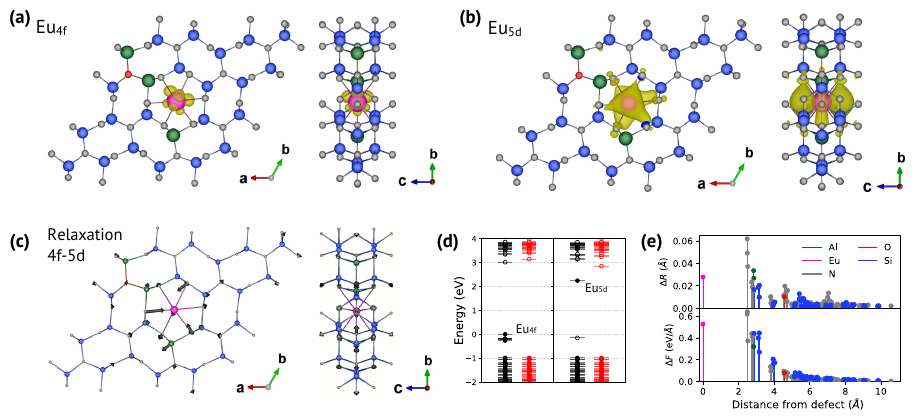}
	\caption{\label{fig:beta_SiAlON_dSCF}
    (a-b) Electronic density associated with the Eu $4f$ and Eu $5d$ states from two points of view.
    (c) Atomic relaxation induced by the electronic transition, scaled by 20 for visualization, (d) Kohn-Sham energy levels computed in a 225-atom supercell at the PBE+U level using the constrained occupation method. 
    Ground state (left): 7 occupied in-gap $4f$ states; excited state (right): occupied $5d$ state. 
    Black (red) lines denote spin-up (down) states. 
    (e) Norm of the atomic displacements $\Delta R$ and forces $\Delta F$ induced by the $5d\to4f$ transition as a function of the distance from the Eu interstitial.
    } 
\end{figure*}

\subsection{Vibronic analysis of the reference structure}
\label{subsec:ref_stru}
The structure is a 2$\times$2$\times$4 supercell of hexagonal $\beta$-Si$_3$N$_4$ (P6$_3$/m), giving a net composition of Si$_{93}$Al$_3$ON$_{127}$:Eu or $z$=0.1875. 
Interestingly, the three dopant species (3Al, 1O, 1Eu) are in the same crystallographic $ab$ plane. 
The Eu lies in a distorted Eu-N$_9$ environment with Eu-N distances varying from 2.48~\AA~to 2.83~\AA.
Figure~\ref{fig:beta_SiAlON_dSCF}(a-b) presents the probability density associated with the highest occupied Kohn-Sham levels in the ground and excited state, and the atomic relaxation (c) induced by the transition (upon 4f-5d absorption), along two views. 
In panel (d), the energy levels are reported in the relaxed ground and excited states. 
Panel (e) shows the decay of the atomic displacements and forces as a function of the distance from the Eu interstitial.  
The highest Eu $4f$-like state detaches itself from the other Eu $4f$ states, a feature rarely observed in Eu-doped phosphors but reported in Ref.~\cite{wang2016elucidating}. 
In the excited state, the Eu $5d$-like orbital shows pronounced $5d_{z^2}$ character along the empty hexagonal channel. 
It is interesting to observe the three-fold-like symmetry of this state, which is rotated by 60$^\circ$ behind the Eu atom. 
A zero-phonon line energy of 2.67~eV is computed, compared with an experimental value of 2.43~eV for low-$z$ $\beta$-SiAlON:Eu$^{2+}$~\cite{takahashi2012origin}.

Regarding the atomic relaxation caused by the electronic transition, the heavy Eu atom is significantly displaced in its Eu-N$_9$ environment, causing a global shift of the Eu-N$_9$ polyhedron in the structure. 
The Eu-N$_9$ polyhedron slightly shrinks upon absorption, but in an anisotropic manner. 
Indeed, only the first three N atoms, contained in the $ab$ Eu-Al-O plane, move toward the Eu atom by around 0.05~\AA.

\begin{figure}[t]
	\centering
	\includegraphics[width=0.95\linewidth]{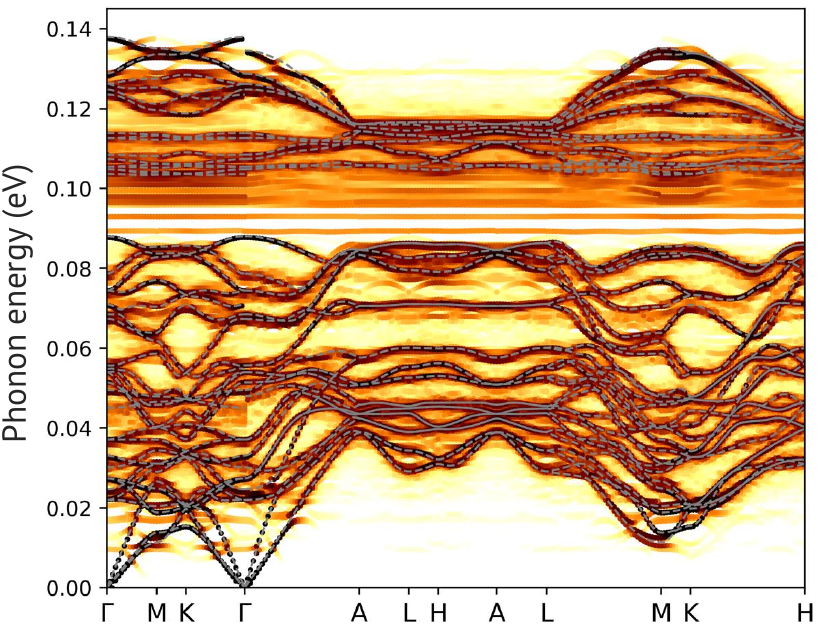}
	\caption{\label{fig:beta_SiAlON_unfolded_bs}
    Unfolded phonon bands of $\beta$-SiAlON:Eu$^{2+}$ along high-symmetry momentum lines of the $\beta$-Si$_3$N$_4$ unit cell. 
    The color scale indicates unfolding weights where darker colors indicate weight closer to unity. 
    The phonon band structure of pristine $\beta$-SiAlON is shown in dashed gray.
    }
\end{figure}

We then use the unfolding technique from Refs.~\cite{allen2013recovering, popescu2012extracting, togo2023implementation} to understand how the pristine phonons of $\beta$-Si$_3$N$_4$ are perturbed by the introduction of the defect complex (one Eu interstitial, three Al and one O substitutions).
Figure~\ref{fig:beta_SiAlON_unfolded_bs} presents the unfolded phonon bands of the 225-atom supercell projected on the high-symmetry $\mathbf{q}$-path of the $\beta$-Si$_3$N$_4$ unit cell. 
The phonon band structure of pristine $\beta$-Si$_3$N$_4$ is shown in dashed gray.
The colored unfolded bands enable the visualization of the broadening of the pristine bands, which reflects slight perturbations of the original phonon modes.
In addition, localized vibrational modes emerge. 
At low frequencies, these modes are predominantly associated with Eu-based phonon vibrations. 
At around 10~meV, we identify a first mode characterized by Eu displacements along the c-axis. 
Around 20~meV, modes involving Eu displacements within the Eu–Al–O plane appear. 
Finally, we observe the formation of modes within the band gap of the pristine $\beta$-Si$_3$N$_4$ phonon spectrum. 
These numerous in-gap modes primarily involve vibrations of the O and Al atoms.

\begin{figure}[t]
	\centering
	\includegraphics[width=0.98\linewidth]{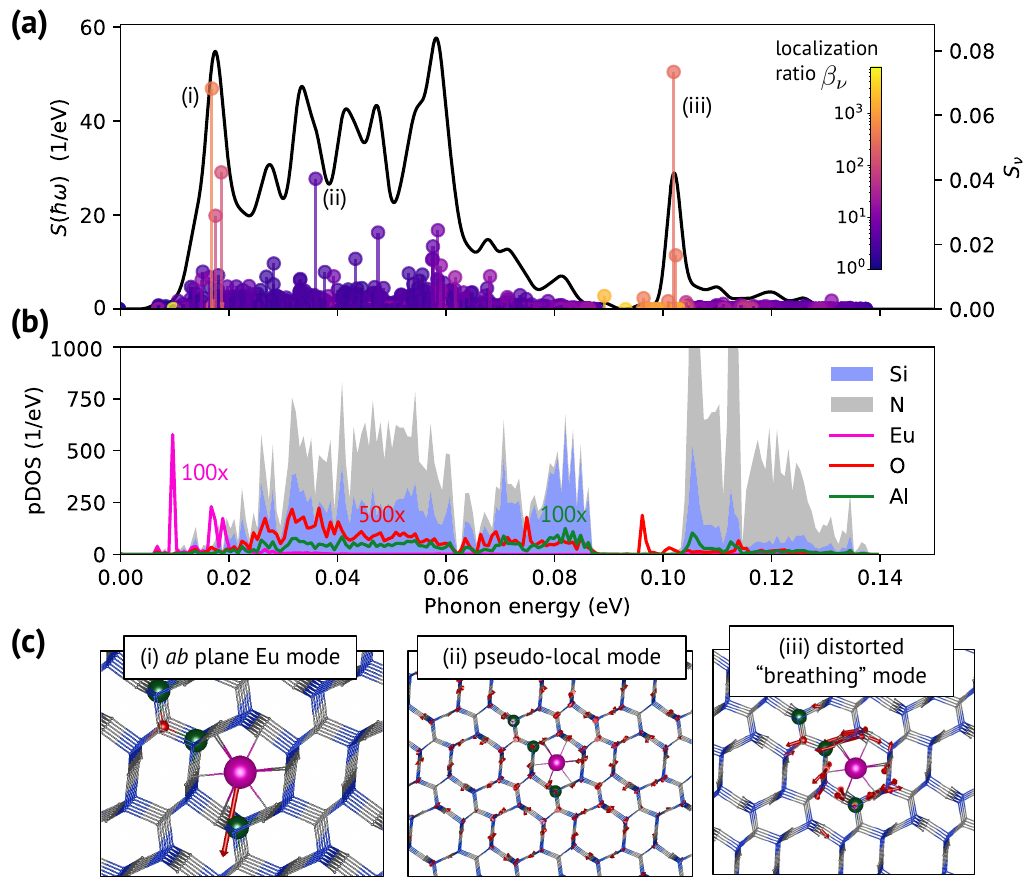}
	\caption{\label{fig:beta_SiAlON_HR_pdos}
    (a) Huang-Rhys spectral function S($\hbar\omega$) (black line) obtained from the individual S$_{\nu}$ (vertical lines) in a 3501-atom supercell, using a 1~meV Gaussian broadening for visualization. 
    Each vertical line is colored according to the localization ratio of the mode $\beta_{\nu}$.
    (b) Atom-projected phonon density of states. 
    (c) Three illustrative high-coupling modes.
    }
\end{figure}

The Huang-Rhys (HR) spectral function computed with a large 5$\times$5$\times$10 supercell containing 3501 atoms is shown in Figure~\ref{fig:beta_SiAlON_HR_pdos}. 
The IFCs embedding procedure~\cite{alkauskas2014, jin2021photoluminescence} was used with a cutoff radius $R_c$=5.7~\AA. 
The individual S$_{\nu}$ are colored according to their localization ratio $\beta_{\nu}$~\cite{alkauskas2014} where $\beta_{\nu}$ $\approx$ 1 represents a bulk-like delocalized mode and $\beta_{\nu}$ $\gg$ 1 corresponds to quasi-local or local modes. 
Three illustrative high-coupling modes are displayed at the bottom of this figure. 
The total HR factor is small ($S$=2.15), indicating a weak coupling regime, for which apparent vibronic peaks are expected. 

The HR spectral function is dominated by a broad feature extending from approximately 20~meV to 60~meV, where the coupling is mainly associated with pristine delocalized vibrational modes and pseudo-local modes (see mode (ii) in Figure~\ref{fig:beta_SiAlON_HR_pdos}). 
Just below 20~meV, a sharp peak is observed, which is characteristic of coupling to Eu-centered local modes moving within the $ab$ plane. 
A further sharp peak at 100~meV is found and is attributed to coupling with below-gap local modes involving O, Si, N, and Al atoms located in the first and second coordination shells of the Eu dopant. 
Visual inspection of the corresponding eigenvector indicates that the displacement pattern qualitatively resembles an asymmetric breathing-type distortion of the Eu–N$_9$ polyhedron.

\begin{figure}[t]
	\centering
	\includegraphics[width=0.95\linewidth]{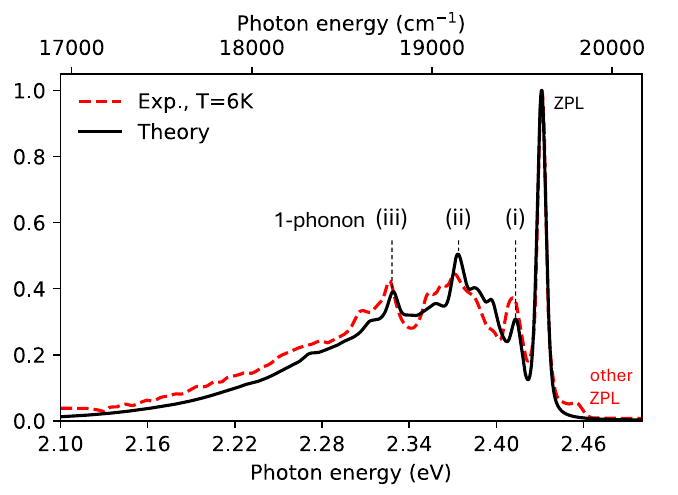}
	\caption{\label{fig:beta_SiAlON_PL_exp}
    Experimental~\cite{takahashi2012origin} and theoretical PL spectra of low-$z$ $\beta$-SiAlON:Eu$^{2+}$ at $T = 6$ K. 
    A rigid energy shift of 0.24~eV was applied to the theoretical spectrum to align the main zero-phonon lines (ZPL). 
    The phonon sidebands includes: (i) an Eu-localized vibrational mode, (ii) broad coupling to delocalized lattice modes, and (iii) distorted breathing modes involving the first and second Eu coordination shells. 
    We ascribe the shoulder at 2.45 eV to a ZPL originating from a different structural variant.
    }
\end{figure}

We then compute PL lineshape at 6~K and compare it to the experiment at the same temperature. 
The ZPL energies are aligned and yield an excellent agreement. 
Temperature-dependent lineshape with experimental comparison is provided in the Supporting information (Fig.~S5).
To understand the microscopic origin of the main features of the PL spectrume, we start with the ZPL line at $\approx$2.41~eV which is the one-phonon contribution of the localized Eu-based mode. 
The next broad peak comes from bulk-like delocalized Si-N modes. 
The last peak at $\approx$2.33~eV is the signature of the one-phonon contribution of distorted breathing modes involving the first and second Eu coordination shells.

Interestingly, we do not reproduce the shoulder at 2.45~eV. 
This shoulder could be due to other structures. 

To validate this hypothesis, we extend our analysis to a systematic study of several structural models at three values of $z$ to evaluate the robustness of the spectroscopic signatures and to identify composition-dependent trends. 
While the selected model shows excellent agreement with the experimental PL lineshape, this agreement does not uniquely resolve the underlying structure, as alternative models may yield comparably good fits. 
This motivates a broader exploration of possible structural scenarios.

\subsection{Vibronic Properties Across Structural and Compositional Variations}

\begin{figure*}[t]
	\centering
	\includegraphics[width=0.9\textwidth]{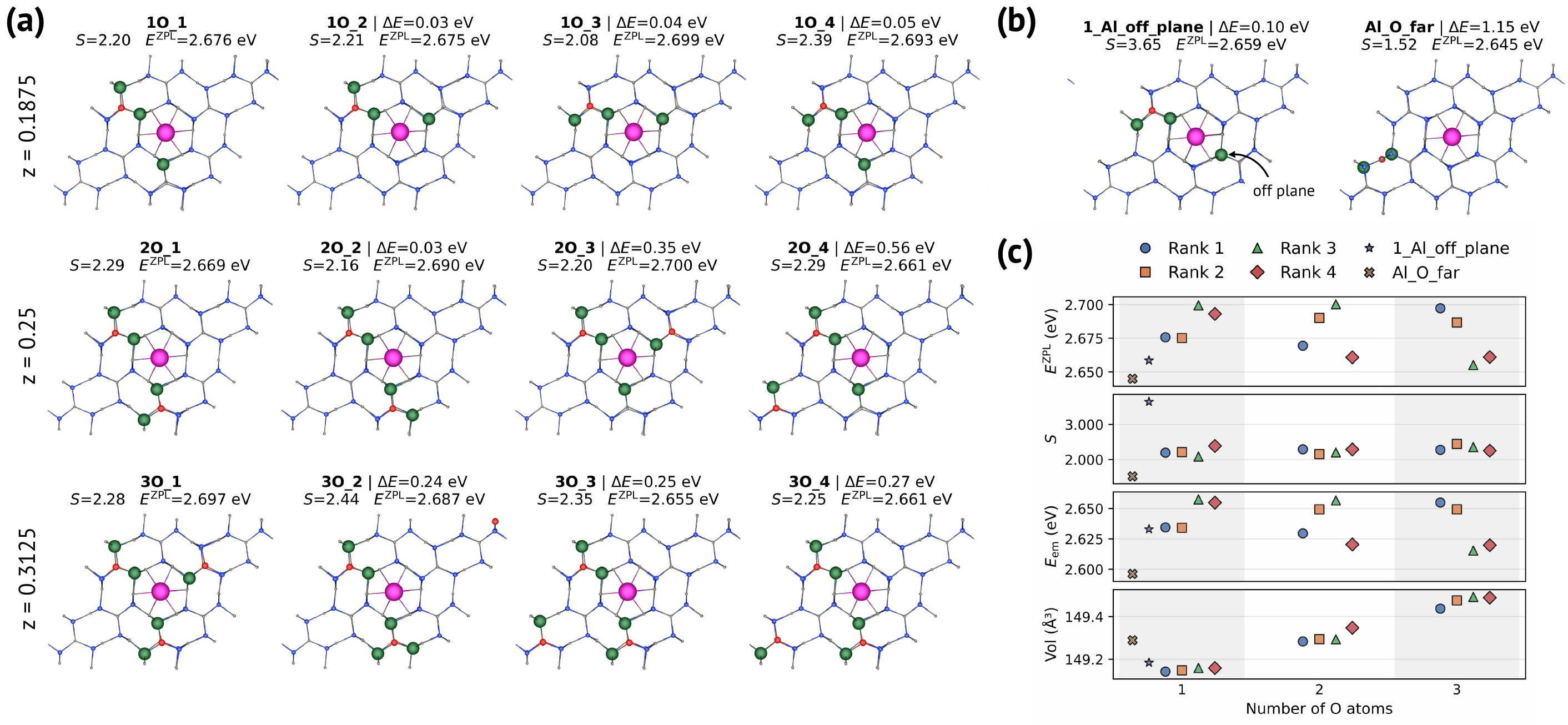}
	\caption{\label{fig:beta_all_structures}
    Overview of the 14 representative structures and their computed properties. 
    (a) Structural models for three compositions ($z$ = 0.1875, 0.25, 0.3125), labeled as nO\_X where n indicates the number of O substitutions and X (1--4) distinguishes different Al/O configurations ordered by increasing energy. 
    Each structure shows the energy difference $\Delta E$ (in eV) relative to the lowest-energy configuration at that $z$, the total Huang-Rhys factor $S$, and the zero-phonon line energy $E^{\rm ZPL}$ (in eV). 
    (b) Two energetically unfavoured structural models: `1\_Al\_off\_plane' with one out-of-plane Al atom and `Al\_O\_far' with a distant Al$_3$O cluster. 
    (c) Scatter plots showing the dependence of computed properties on the number of O substitutions: ZPL energy, total Huang-Rhys factor $S$, estimated emission energy, and unit cell volume. 
    The rank number in the legend corresponds to the configuration label X.
    }
\end{figure*}

We compute the photoluminescence properties of 14 representative structural models spanning three compositions ($z$ = 0.1875, 0.25, 0.3125) with four distinct Al/O configurations per composition, as described in Sec.~\ref{Sec:method}. 
Figure~\ref{fig:beta_all_structures}(a) shows an overview of these structures. The energetically favoured are labeled as nO\_X where n indicates the number of O substitutions (1, 2, or 3) and X (1--4) distinguishes different Al/O configurations at fixed $z$, ordered by increasing energy from left to right. The two representative unfavoured configurations are also displayed in Figure~\ref{fig:beta_all_structures}(b) : `1\_Al\_off\_plane' has one Al atom outside the $ab$ plane of the Eu interstitial, while `Al\_O\_far' has an Al$_3$O cluster located far from the Eu site.
The atomic sphere radii of Si and N are reduced, whereas those of the dopant species are enlarged to accentuate the distinct configurational arrangements.
Each structure is shown with its energy difference $\Delta E$ relative to the lowest-energy configuration at that $z$ value, along with the computed Huang-Rhys factor $S$ and zero-phonon line energy $E^{\rm ZPL}$. 
The reference structure analyzed in detail in the previous section corresponds to 1O\_1. 

Figure~\ref{fig:beta_all_structures}(c) shows scatter plots of computed properties (ZPL energy, total Huang-Rhys factor $S$, estimated emission energy, and unit cell volume) as a function of the number of O substitutions. 
Luminescence properties are computed using a one-dimensional configuration coordinate model~\cite{bouquiaux2021importance, jia2017first}.

Several key trends emerge from this exploration. 
First, all low-energy configurations have Al and O atoms located in the same crystallographic $ab$-plane as the Eu interstitial, consistent with the Wang \textit{et al.} structural rules~\cite{wang2016elucidating}. 
Focusing on this class of energetically favoured configurations, the ZPL energy shows weak sensitivity to Al/O configuration at fixed $z$, with small variations within each composition (see rows in Figure~\ref{fig:beta_all_structures}). 
For the $z$ = 0.1875 composition (1 O, 3 Al), we observe a 25~meV variation in ZPL energy between the two lowest-energy configurations (1O\_1 and 1O\_2) and the other two (1O\_3 and 1O\_4). 
This difference arises from the distinct arrangement of the OAl$_2$ cluster near the Eu atom (visible in the left region of the structural models). 
The resulting 25~meV shift (or $\sim$200 cm$^{-1}$) may explain the small shoulder observed in the experimental PL spectrum around 2.45~eV, identified as an 'other ZPL' feature in Figure~\ref{fig:beta_SiAlON_PL_exp} ~\cite{takahashi2012origin, zhang2025narrow}, and that we reproduce in Fig.~S7 of the Supporting Information.

At higher O content, all ZPL energies remain equal to or below that of the lowest-energy $z$=0.1875 configuration, indicating that increasing Al/O substitution systematically red-shifts or maintains the ZPL energy without producing blue-shifts. 
Moreover, the Huang-Rhys factor $S$ increases modestly with $z$. 
These combined trends produce a red-shift in the estimated emission energy with increasing $z$ (Figure~\ref{fig:beta_all_structures}(c), bottom left), consistent with experimental observations~\cite{takahashi2012origin, zhang2017controlling, zhang2025narrow}. 
The unit cell volume (Figure~\ref{fig:beta_all_structures}(c), bottom right) increases systematically with $z$, in agreement with experimental trends~\cite{zhang2025narrow}.
Additional structural properties for all 14 configurations, including Eu-N first-shell distances and second-shell compositions, are reported in Section S4 of the Supporting Information.

\begin{figure}[t]
	\centering
	\includegraphics[width=0.95\linewidth]{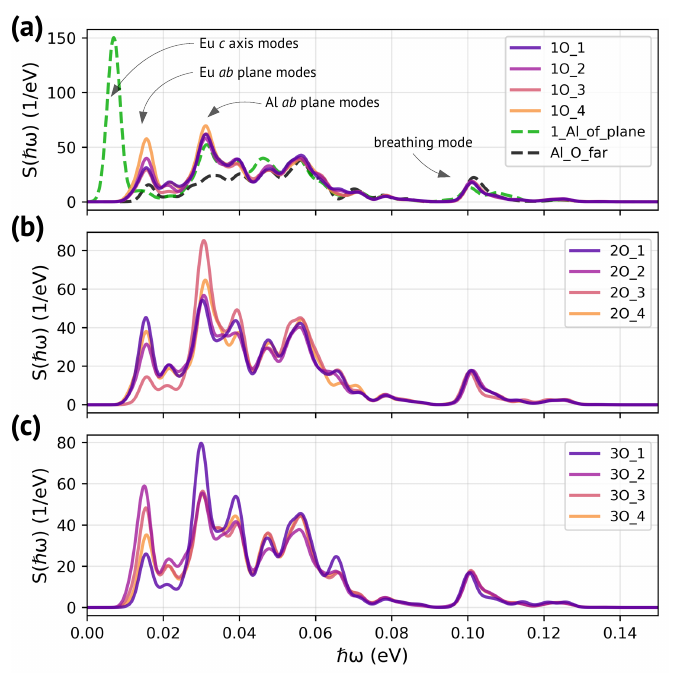}
	\caption{\label{fig:beta_all_Shw}
    Huang-Rhys spectral functions 
    $S(\hbar\omega)$
     for all 14 representative structures (solid lines), showing remarkable similarity in shape across compositions and Al/O configurations. 
    (a-b-c) panels shows the spectral functions with (1-2-3) oxygen atoms in the supercell structure. 
    The energetically unfavoured structures `1\_Al\_off\_plane' (green dashed line) and `Al\_O\_far' (black dashed line) exhibit distinct spectral features, indicating stronger and weaker coupling to Eu-localized modes, respectively. 
    }
\end{figure}

Figure~\ref{fig:beta_all_Shw} compares the Huang-Rhys spectral functions $S(\hbar\omega)$ across all 14 structures.
Remarkably, $S(\hbar\omega)$ exhibits a nearly identical shape in compositions and Al/O configurations for the 12 energetically favoured structural models.
The characteristic spectral shape persists regardless of Al/O arrangement or $z$ value. 
Only the intensities of Eu-based mode peaks vary significantly, reflecting subtle changes in Eu displacements during the electronic transition induced by different local Al/O environments. 
For example, 2O\_3 shows reduced coupling to Eu-localized modes because the Al three-fold symmetry around Eu constrains Eu displacements during the
electronic transition.
The integrated spectral function (total Huang-Rhys factor $S$) increases slightly with $z$, as discussed above. 
Computed PL lineshapes for all 14 configurations at $T = 6$~K, including Boltzmann-weighted spectra for the $z = 0.1875$ composition, are shown in Section S4 of the Supporting Information.

This robustness has important implications. 
Weak electron-phonon coupling ($S \approx 2.15$) is intrinsic to energetically favoured Al-O configurations around the Eu-N$_9$ site, not a consequence of specific Al/O ordering or low $z$ values. 
This explains why the resolved vibronic structure persists even at room temperature and relatively high $z$ values. 
However, this robustness is restricted to energetically-favoured configurations with Al and O atoms in the same $ab$-plane as the Eu interstitial.

To test these limits, we computed vibronic properties for deliberately energetically unfavoured configurations: structures with out-of-plane Al placement and with Al/O distant from the Eu site. 
Out-of-plane Al/O substitutions exhibit significantly higher Huang-Rhys factors, with $S(\hbar\omega)$ showing strong coupling to Eu-localized modes involving $c$-axis displacement into the empty hexagonal channel (green dashed curve, Fig.~\ref{fig:beta_all_Shw}). 
These substitutions induce Eu-N$_9$ polyhedron relaxations with significant $c$-axis components, strengthening coupling to Eu-localized modes along this direction. 
While rare at low $z$, such energetically unfavoured configurations may become more prevalent at higher $z$, potentially contributing to the experimentally observed increase in $S$ with $z$.
In contrast, configurations with distant Al/O clusters (black curve, Figure~\ref{fig:beta_all_Shw}) show reduced coupling to Eu-based modes because the Eu-N$_9$ environment remains less perturbed.
Al-O modes at $\sim$30 meV are likewise weakened, since the remote Al-O cluster experiences minimal perturbation from the electronic transition. 
This model fails to reproduce the experimental vibronic peaks, missing the characteristic Eu phonon replica peak. 
Only energetically favoured ab-planar configurations simultaneously achieve moderate coupling strength and match the experimental spectral fingerprints.

\subsection{Implications for Experimental Spectra Interpretation}
\label{subsec:implications}

Our combined results provide a comprehensive picture for interpreting the experimental photoluminescence spectra of $\beta$-SiAlON:Eu$^{2+}$ across varying $z$ values. 
First, the excellent agreement between the computed and experimental 6~K PL lineshapes (Sec.~\ref{subsec:ref_stru}) validates the Eu-N$_9$ coordination model with in-plane Al/O coordination through detailed vibronic analysis. 
The characteristic phonon replicas (Eu-localized modes at $\sim$20~meV, bulk Si-N modes at 30--60~meV, and breathing-type distortions at $\sim$100~meV) are accurately reproduced, establishing these vibronic peaks as a structural signature of the Eu-N$_9$ environment. 
Second, experiments show that the same dominant ZPL persists from $z \approx 0.03$ to $z > 0.3$~\cite{zhang2025narrow}. 
The Monte Carlo sampling identifies the Eu–N$_9$ site with in-plane 3Al+1O coordination as the energetically favoured local structure across all studied compositions, stabilized by local charge compensation.
This persistence arises because $z$ controls the Al/O content of the host lattice rather than the Eu dopant concentration. 
Each interstitial Eu$^{2+}$ introduces a local charge imbalance, which drives nearby Al and O atoms to cluster around the Eu site irrespective of the bulk $z$. 
At low $z \approx 0.03$, they preferentially aggregate near Eu, predominantly forming the reference 3Al+1O in-plane coordination.

Our calculations reveal a small number of energetically accessible variants (four structures within 50 meV at $z = 0.1875$), all of which produce very similar vibronic signatures. 
Experimentally, the reference structure dominates the PL spectrum, while minor contributions from these variants appear as weak shoulders.
At higher $z$, the increased Al/O concentration enables greater configurational diversity. 
While the reference 3Al+1O motif remains present, other ab-planar and energetically unfavoured configurations (such as out-of-plane Al coordination) become increasingly populated. 
These additional environments exhibit ZPL energies comparable to or lower than that of the reference structure, resulting in a red shift of the emission and progressive spectral broadening due to both a distribution of ZPL energies and broader individual vibronic peaks.

\section{Conclusions}

We studied the long-standing question of the precise atomic-scale structure around Eu$^{2+}$ activators in $\beta$-SiAlON:Eu$^{2+}$ through first-principles vibronic spectroscopy. 
The excellent agreement between the computed photoluminescence spectrum and experimental data at 6~K, reproducing the characteristic vibronic peaks with correct energies and relative intensities, validates the Eu-N$_9$ coordination model with Al, O, and Eu confined to the same crystallographic $ab$-plane. 
We investigated 14 representative structures (12 ab-planar and 2 out-of-plane configurations) to understand the physical origin of this material's unusual optical properties. 
The characteristic vibronic peaks (Eu-localized modes at $\sim$20~meV, bulk Si-N vibrations at 30--60~meV, and breathing-type distortions at $\sim$100~meV) directly reflect the specific geometry of in-plane Al/O coordination around the Eu-N$_9$ polyhedron. 
The weak electron-phonon coupling ($S \approx 2.15$) is intrinsic to energetically favourable in-plane configurations and remains remarkably robust across different Al/O arrangements. 
This explains why $\beta$-SiAlON:Eu$^{2+}$ retains resolved vibronic features even at room temperature and narrow emission characteristics relatively insensitive to microscopic synthesis variations, critical properties for commercial applications.
A key insight concerns local charge compensation in governing dopant coordination. 
Each interstitial Eu$^{2+}$ drives preferential aggregation of nearby Al and O atoms, predominantly forming 3Al+1O in-plane coordination regardless of bulk composition $z$. 
This mechanism explains the persistence of the dominant zero-phonon line from $z \approx 0.03$ to $z > 0.3$. 
At low $z$, this configuration dominates, producing narrow emission with well-resolved vibronic peaks. 
At higher $z$, increased configurational diversity produces the experimentally observed red-shift and spectral broadening while maintaining the underlying structural motif.
By examining energetically unfavoured configurations, we establish the structural requirements for favorable optical properties. 
Out-of-plane Al substitutions significantly increase electron-phonon coupling, while distant Al/O clusters fail to reproduce the experimental vibronic peaks.
These configurations, though rare at low $z$, may contribute to spectral broadening at higher compositions.
The computational framework developed here, integrating Monte Carlo sampling, machine-learned potentials, $\Delta$SCF calculations, and IFCs embedding for supercells up to 3501 atoms, provides a general methodology for investigating doped phosphor systems where experimental structural characterization remains challenging. 
%

% Supporting
\medskip
\textbf{Supporting Information} \par
Section S1 details the Monte Carlo sampling methodology. Section S2 presents the convergence of the Huang–Rhys spectral function $S(\hbar\omega)$ with supercell size, along with a direct comparison between DFT and \textsc{MatterSim} interatomic force constants for the reference structure. Section S3 shows the temperature-dependent photoluminescence lineshapes for the reference structure (1O\_1) compared with experimental spectra. Section S4 reports full structural and luminescent properties for all 14 configurations, computed PL lineshapes at 6~K, and Boltzmann-weighted spectra. 

The data associated with this work are publicly available on the Materials Cloud archive at \href{https://doi.org/10.24435/materialscloud:4a-g6}{https://doi.org/10.24435/materialscloud:4a-g6}. 
It contains the relaxed ground-state structures, \textsc{Phonopy} force constants, \textsc{ABINIT} $\Delta$SCF outputs, and displacement visualization 
files for all 14 configurations studied. 
Computed emission properties  for all structures are provided. 
An interactive viewer of the phonon modes for the reference structure is also provided.

% Acknowledgements
\medskip
\textbf{Acknowledgments} \par
J. B. acknowledge support from the F.R.S.-FNRS. 
S. P. is a Research Associate of the Fonds de la Recherche Scientifique - FNRS.
This work was supported by the Fonds de la Recherche Scientifique - FNRS under Grants number T.0183.23 (PDR) and  T.W011.23 (PDR-WEAVE). 
This publication was supported by the Walloon Region in the strategic axe FRFS-WEL-T.
Y. J. acknowledges funding support from National Key R\&D Program (No.2022YFB3503800), Natural Science Foundation of Hebei Province (No.E2021203126) and Cultivation Project for Basic Research and Innovation of Yanshan University (No.2021LGQN033).
Computational resources have been provided by the supercomputing facilities of the Université catholique de Louvain (CISM/UCL) and the Consortium des Équipements de Calcul Intensif en Fédération Wallonie Bruxelles (CÉCI) funded by the Fonds de la Recherche Scientifique de Belgique (F.R.S.-FNRS) under convention 2.5020.11 and by the Walloon Region, as well as by the Tier-1 supercomputer of the Walloon Region (Lucia) with infrastructure funded by the Walloon Region under the grant agreement n°1910247.

% References
\medskip
\bibliography{main} 

\end{document}

\title [mode = title]{SI: Micro-environment of the Eu interstitial in the $\beta$-SiAlON:Eu$^{2+}$ green phosphor}

\author{Julien Bouquiaux}
\email{julien.bouquiaux@matgenix.com}
\affiliation{Institute of Condensed Matter and Nanosciences, Universit\'{e} catholique de Louvain, Chemin des \'{e}toiles 8, bte L07.03.01, B-1348 Louvain-la-Neuve, Belgium}
\affiliation{Matgenix, A6K Advanced Engineering Centre, B-6000 Charleroi, Belgium}
\author{Samuel Ponc\'{e}}
\affiliation{Institute of Condensed Matter and Nanosciences, Universit\'{e} catholique de Louvain, Chemin des \'{e}toiles 8, bte L07.03.01, B-1348 Louvain-la-Neuve, Belgium}
\affiliation{WEL Research Institute, avenue Pasteur 6, 1300 Wavre, Belgium}
\author{Yongchao Jia}
\affiliation{Yanshan University,Hebei Key Laboratory of Applied Chemistry, Yanshan University, Hebei Street 438, 066004, Qinhuangdao, P. R. China}
\author{Masayoshi Mikami}
\affiliation{Analysis Technology Laboratory, Science $\&$ Innovation Center, Mitsubishi Chemical Corporation, 1000,
Kamoshida-cho Aoba-ku, Yokohama, 227-8502, Japan}
\author{Xavier Gonze}
\affiliation{Institute of Condensed Matter and Nanosciences, Universit\'{e} catholique de Louvain, Chemin des \'{e}toiles 8, bte L07.03.01, B-1348 Louvain-la-Neuve, Belgium}

\maketitle

%\tableofcontents
%%%%%%%%%%%%%%%%%%%%%%%%%%%%%%%%%%%%%%%%%%%%
\section{Monte-Carlo exploration}
%%%%%%%%%%%%%%%%%%%%%%%%%%%%%%%%%%%%%%%%%%%%
\begin{figure}[t]
    \centering
    \includegraphics[width=0.9\linewidth]{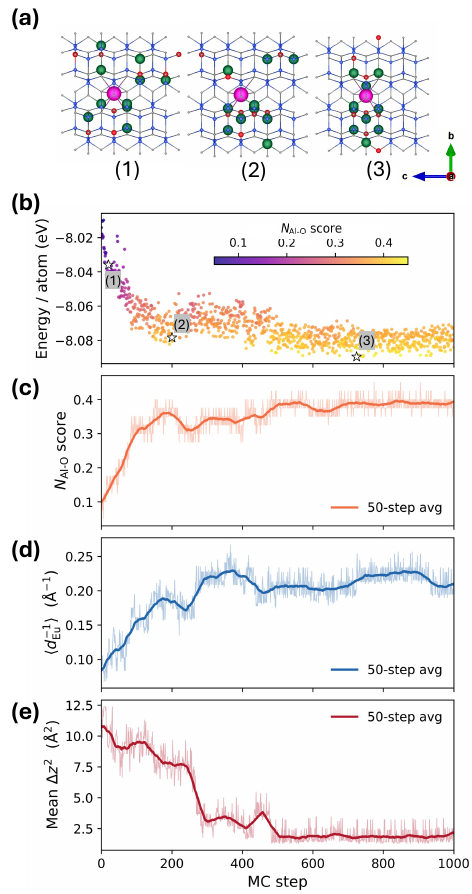}
    \caption{\label{fig:SI_mc_chain_example}
    Spontaneous Al--O--Eu aggregation in an unbiased MC chain ($z \approx 0.625$, 10\,Al, 8\,O, 1\,Eu, 225-atom supercell).
    \textbf{(a)}~Structure snapshots marked by stars in panel~(b): (1)~step~20 (random), (2)~step~200 (partial clustering), (3)~lowest-energy configuration visited.
    Color code: Eu (magenta), Al (green), O (red), N (grey), Si (blue).
    \textbf{(b)}~Energy per atom along the chain, colored by the Al--O aggregation score $N_\mathrm{Al\text{-}O}$ (fraction of possible Al--O nearest-neighbor bonds realized within 3\,\AA; 0: Al and O fully separated, 1: every Al fully coordinated by O).
    \textbf{(c--e)}~Evolution of the three bias-equation descriptors: Al--O aggregation score, Eu-proximity score $\langle d_\mathrm{Eu}^{-1}\rangle$ (mean inverse distance of Al/O to Eu), and mean squared $z$-displacement $\Delta z^2$ of Al/O relative to Eu. Thick lines are 50-step running averages.
    }
\end{figure}

Structural models for $\beta$-Si$_{6-z}$Al$_z$O$_z$N$_{8-z}$:Eu are generated via Monte Carlo (MC) sampling of the Al/O substitution space using the \textsc{MACE-MP-0a (large)} machine-learned interatomic potential~\cite{batatia2022mace, batatia2025design} for fast energy evaluation.
Starting from a $2\times2\times4$ supercell of hexagonal $\beta$-Si$_3$N$_4$ (224 Si/N atoms) with one Eu, random Al$\to$Si and O$\to$N substitutions are first performed to reach the target composition. 
At each MC step, an attempted move randomly exchanges either an Al/Si pair or an O/N pair, accepted with the Metropolis criterion:
\begin{equation}
    P^\mathrm{accept} = \min\!\left(1,\, e^{-\Delta E / k_B T}\right),
\end{equation}
where $\Delta E$ is the total energy change per supercell evaluated by MACE. 
A simulated annealing decreases the temperature linearly from $T^\mathrm{start}=0.5$~eV to $T^\mathrm{end}=0.001$~eV over \num{50000} steps. 
No structural relaxation is performed at each step.
We validate this by comparing the result of 200 without relaxation and 200 with MACE-relaxed energies and find that the 
MACE unrelaxed energies rank configuration correctly with a Spearman rank correlation of $\rho = 0.974$.

Exploratory runs reveal that low-energy structures consistently feature Al--O nearest-neighbor clustering, proximity of both dopants to the Eu interstitial, and confinement of Al/O to the same (001) $ab$-plane as Eu. 
To improve sampling efficiency, we implement a biased move-proposal scheme 
that weights swap candidates as
\begin{equation}
\label{eq:weights}
    w \sim \exp\!\left(\alpha N_{\mathrm{Al\text{-}O}} + \beta \cdot d_{\mathrm{Eu}}^{-1}\right) \cdot \exp\!\left(-\gamma \Delta z^{2}\right),
\end{equation}
where $N_{\mathrm{Al\text{-}O}}$ counts Al--O nearest-neighbor pairs, 
$d_{\mathrm{Eu}}$ is the distance to the Eu site, and $\Delta z$ is the 
$z$-coordinate difference relative to Eu.
This equation was derived heuristically to reflect the structural 
motifs identified in the exploratory runs: the first term favors candidates 
that increase Al--O nearest-neighbor contact and proximity to Eu, while the 
second term penalizes moves that displace Al or O 
away from the Eu $(001)$ plane.
Eight independent chains with different random seeds are run per composition, starting from different random initial structures. 

For each target composition ($z = 0.1875$, $0.25$, $0.3125$), corresponding to (3\,Al, 1\,O), (4\,Al, 2\,O), and (5\,Al, 3\,O) substitutions respectively, the MC sampling consistently yields low-energy structures in which Al, O, and Eu are co-confined to the same (001) $ab$-plane, in line with Ref.~\cite{wang2016elucidating}. 
We additionally explore the energetics of displacing the Eu atom along $z$, finding a well-defined energy minimum at the (001) plane occupied by the Al/O dopants.

After MC sampling, the four lowest-energy distinct configurations per composition are selected and relaxed with DFT, yielding 12 structural models. 
Two additional energetically-unfavoured configurations (one with an out-of-plane Al atom and one with the Al$_3$O cluster far from Eu) are also included to assess the sensitivity of the vibronic properties to the Al/O arrangement.

As an illustrative example, Fig.~\ref{fig:SI_mc_chain_example} shows a single unbiased MC chain at high composition ($z \approx 0.625$, 10\,Al, 8\,O, 1\,Eu in a 225-atom supercell) under simulated annealing ($T$: $0.5 \to 0.001$~eV over 1000 steps), starting from a random substitution configuration.
Three representative snapshots are shown at the top: structure~(1) at step~20 (disordered), structure~(2) at step~200 (partial clustering), and structure~(3) at the lowest-energy visited configuration.
The four panels track the energy per atom and the three structural descriptors entering the bias weight $w$ (Eq.~\ref{eq:weights}): the Al--O aggregation score $N_\mathrm{Al\text{-}O}$, the Eu-proximity score $\langle d_\mathrm{Eu}^{-1}\rangle$, and the mean out-of-plane deviation $\Delta z^2$ of Al/O relative to Eu.
All three motifs self-organise spontaneously, confirming that Al--O--Eu aggregation and coplanar confinement are intrinsic features of the energy landscape rather than artifacts of the production bias.
At such high $z$, full confinement to a single (001) plane is geometrically unlikely given the large number of dopants.

%%%%%%%%%%%%%%%%%%%%%%%%%%%%%%%%%%%%%%%%%%%%
\section{Huang-Rhys spectral function}
\label{sec:SI_HR}
%%%%%%%%%%%%%%%%%%%%%%%%%%%%%%%%%%%%%%%%%%%%

%\subsection{Convergence with supercell size}
%\label{subsec:SI_HR_convergence}

\begin{figure}[t]
    \centering
    \includegraphics[width=0.98\linewidth]{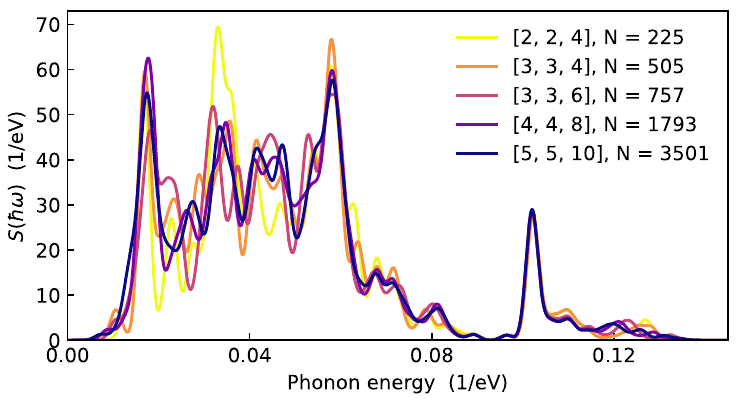}  
    \caption{Convergence of the Huang-Rhys spectral function $S(\hbar\omega)$ with respect to supercell size for the reference structure, where $N$ is the number of atoms. }
    \label{fig:SI_HR_convergence}
\end{figure}

\begin{figure}[t]
    \centering
    \includegraphics[width=0.6\linewidth]{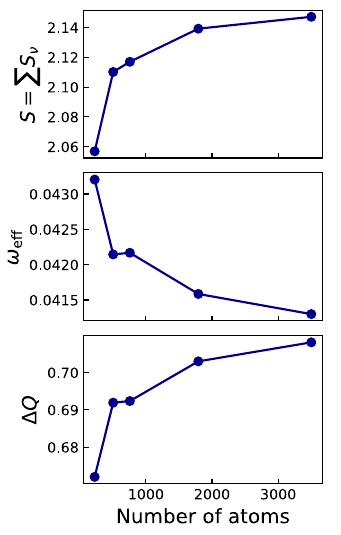}  
    \caption{Convergence of the total Huang-Rhys factor $S$, the effective phonon frequency $\omega_{\rm{eff}}$, and the mass-weighted total displacement $\Delta Q$.}
    \label{fig:SI_HR_convergence_2}
\end{figure}

The Huang-Rhys spectral function $S(\hbar\omega)$ is computed using the IFCs embedding procedure~\cite{alkauskas2014, jin2021photoluminescence,razinkovas2021vibrational,bouquiaux2026lumabi} with a cutoff radius $R_c = 5.7$~\AA. 
To assess convergence with respect to the supercell size, we computed $S(\hbar\omega)$ for several embedding supercells.

\begin{figure}[t]
    \centering
    \includegraphics[width=0.95\linewidth]{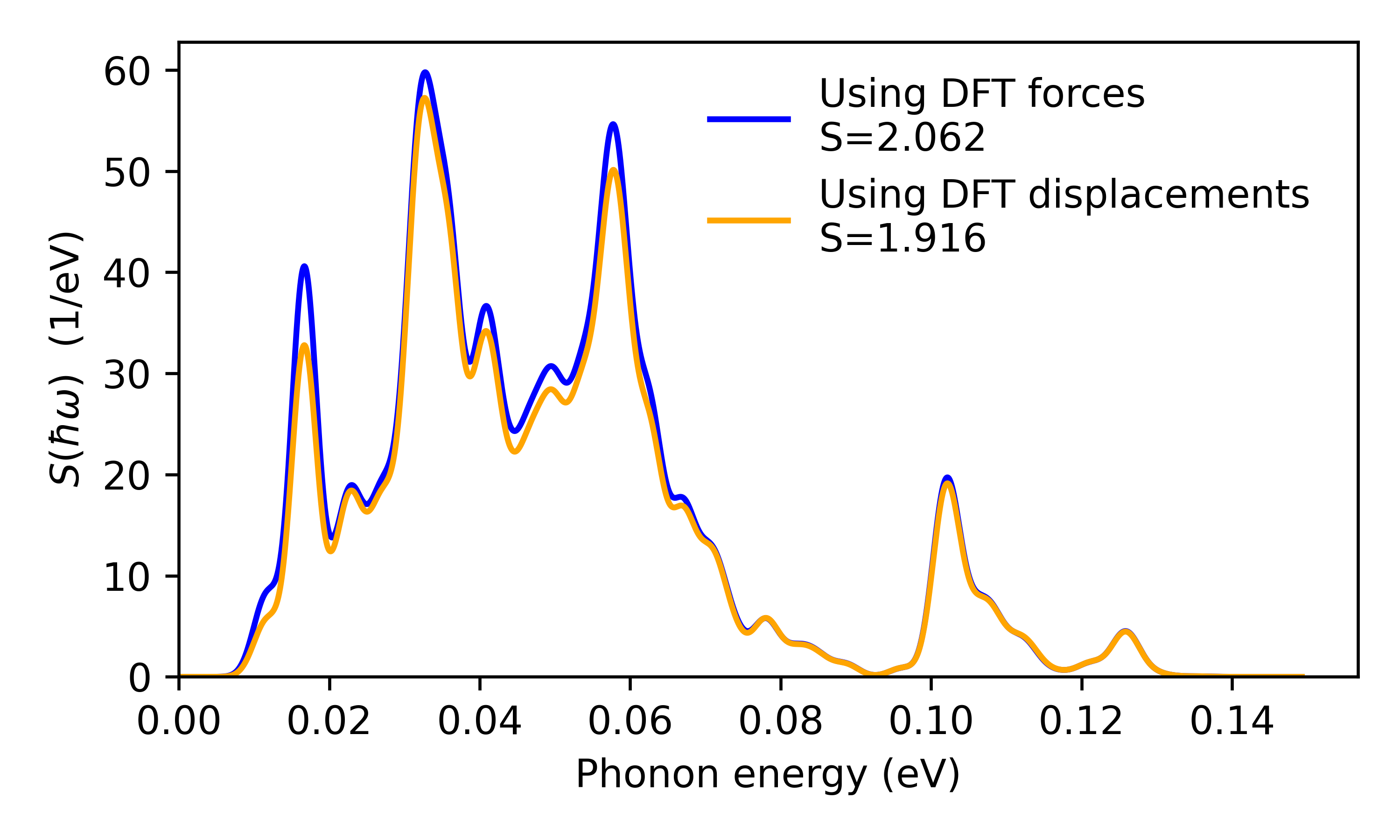}  
    \caption{Comparison of the Huang-Rhys spectral function $S(\hbar\omega)$ obtained with the forces and the displacements.}
    \label{fig:force_dis}
\end{figure}

Figures~\ref{fig:SI_HR_convergence} and \ref{fig:SI_HR_convergence_2} shows $S(\hbar\omega)$ for the reference structure (1O\_1) computed with supercells of increasing size up to 3501 atoms.
The non-monotonic behavior between the second and the third data-point in Figure~\ref{fig:SI_HR_convergence_2} comes from a change in the supercell size ratio ($a/c$). 
When using the defect phonon in the 225-atom supercell, the use of the displacements gives $S$=1.92 and $S$=2.06 for the forces, see Fig.~\ref{fig:force_dis}, indicating a sufficient validity of the harmonic approximation in this system, and ensuring that the use of forces in large supercells is well justified~\cite{jin2021photoluminescence, bouquiaux2023first}.

\begin{figure}[t]
    \centering
    \includegraphics[width=0.98\linewidth]{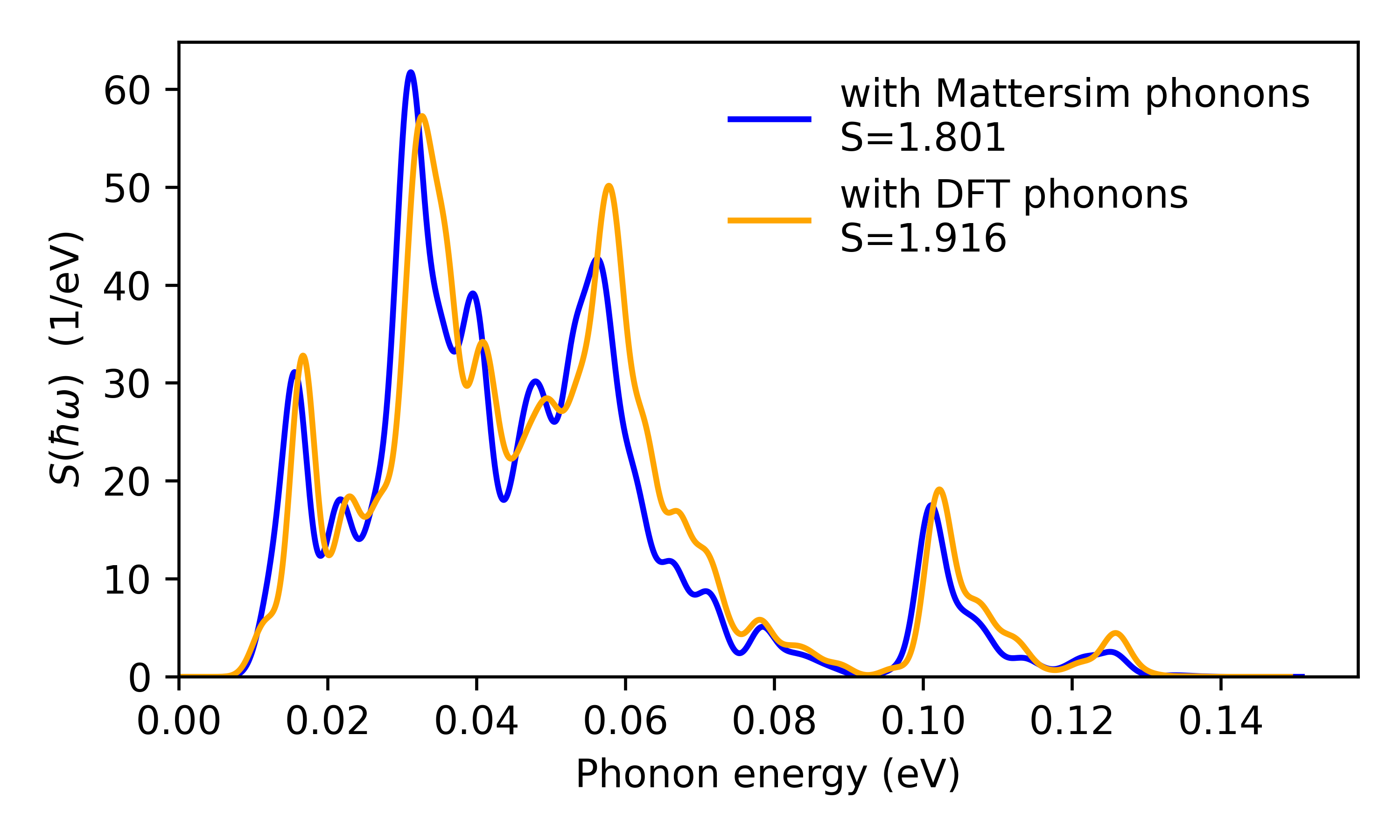}
    \caption{Comparison of the Huang-Rhys spectral function $S(\hbar\omega)$ computed with DFT and machine-learned interatomic potential IFCs for the reference structure.}
    \label{fig:SI_HR_DFT_MLIP}
\end{figure}

We also validate the use of the machine-learned IFCs~\cite{yang2024mattersim} for the reference structure by comparing to DFT results.
Figure~\ref{fig:SI_HR_DFT_MLIP} compares the resulting $S(\hbar\omega)$ obtained from the two approaches. 
The good agreement validates the use of \textsc{Mattersim} IFCs for the remaining structural models studied in this work.

\begin{figure}[t]
    \centering
    \includegraphics[width=0.95\linewidth]{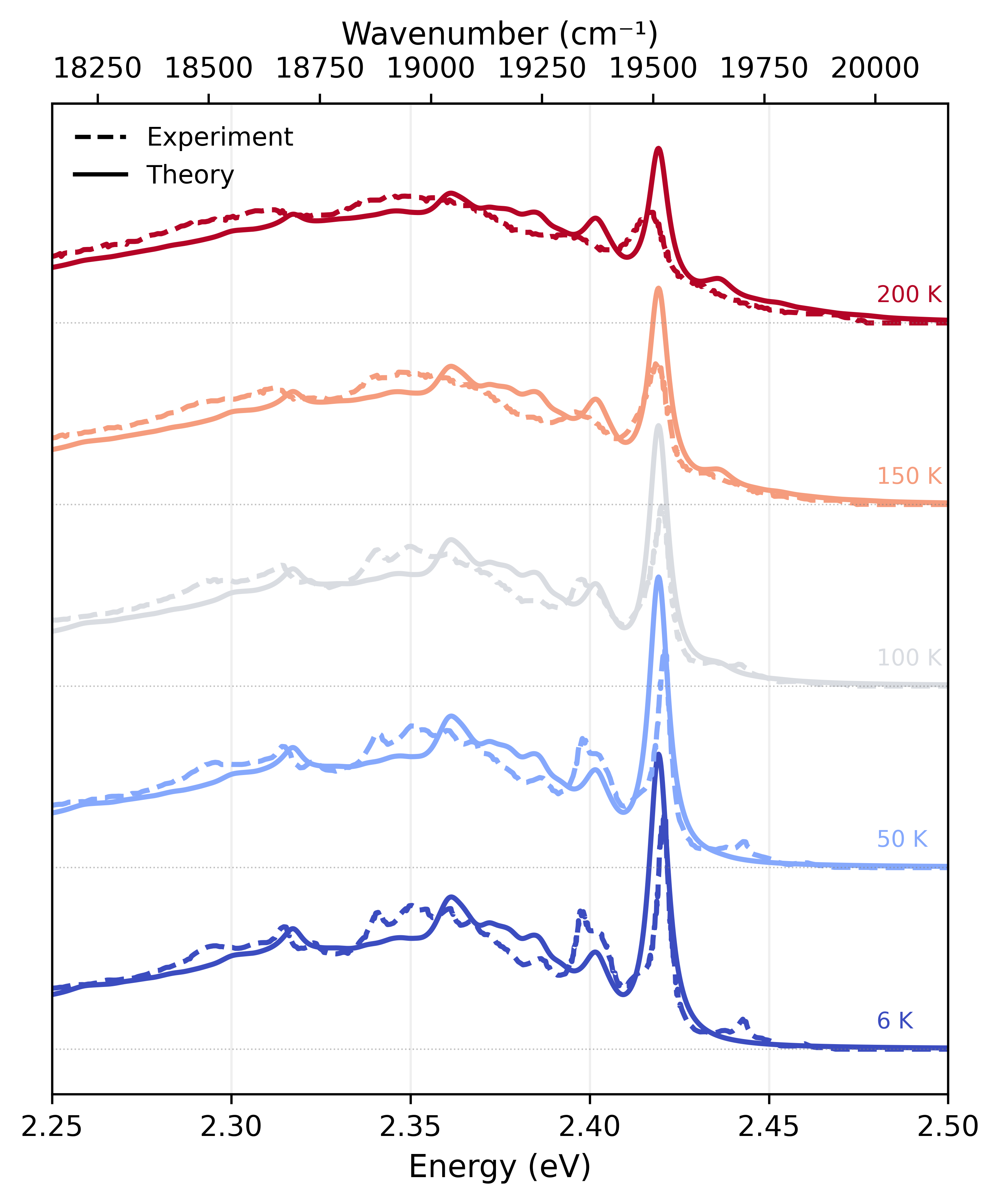} 
    \caption{Computed photoluminescence lineshapes for the reference structure (1O\_1) at several temperatures, compared with experimental lineshapes from reference~\cite{zhang2025narrow}. }
    \label{fig:SI_Tdep}
\end{figure}

%%%%%%%%%%%%%%%%%%%%%%%%%%%%%%%%%%%%%%%%%%%%
\section{Temperature-dependent PL lineshapes for the reference structure}
\label{sec:SI_Tdep}
%%%%%%%%%%%%%%%%%%%%%%%%%%%%%%%%%%%%%%%%%%%%

Figure~\ref{fig:SI_Tdep} shows the computed photoluminescence lineshapes for the reference structure (1O\_1) at several temperatures. 
The lineshape is computed via the generating function approach of Eqs.~(2)--(3) of the main text with the largest supercell (3501 atoms), and are compared with the experimental spectra from reference~\cite{zhang2025narrow}. 
The ZPL energies are aligned.

%%%%%%%%%%%%%%%%%%%%%%%%%%%%%%%%%%%%%%%%%%%%
\section{Structural models}
\label{sec:SI_structures}
%%%%%%%%%%%%%%%%%%%%%%%%%%%%%%%%%%%%%%%%%%%%

\begin{figure}[t]
    \centering
    \includegraphics[width=0.99\linewidth]{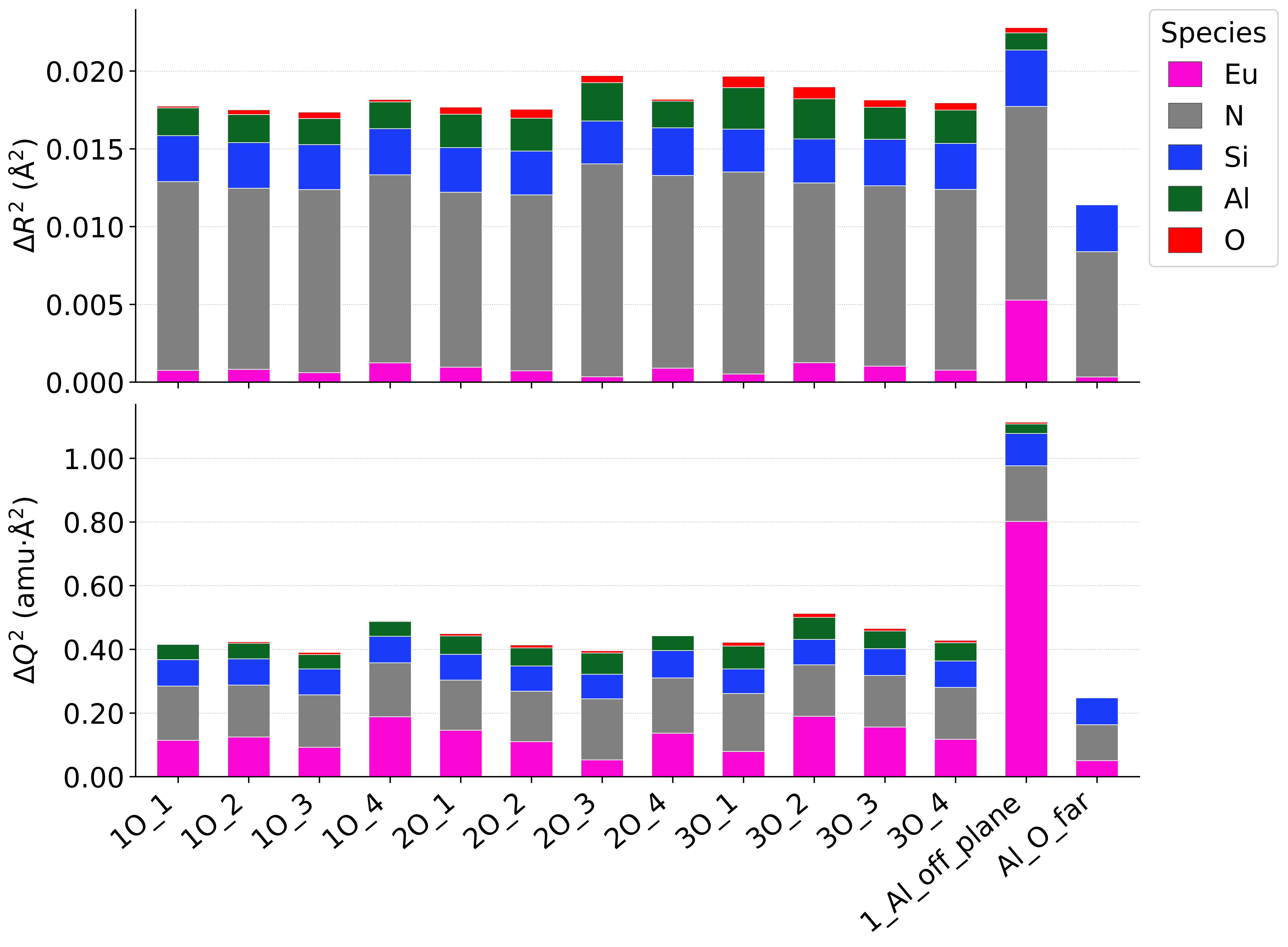}
    \caption{\label{fig:delta_Q2_species}
        Species-resolved structural relaxation upon the $5d \leftarrow 4f$ optical transition of Eu$^{2+}$ for all configurations
        considered in this work.
        }
\end{figure}

Figure~\ref{fig:delta_Q2_species} presents the species-resolved structural relaxation upon the $5d \rightarrow 4f$ transition of Eu$^{2+}$ for 
all configurations considered in this work. 
The top panel shows the total atomic relaxation,
\begin{equation}
    (\Delta R)^2 = \sum_{\alpha i}\left(R^{\mathrm{exc}}_{\alpha i}-R^{\mathrm{GS}}_{\alpha i}\right)^2,
\end{equation}
and the bottom panel shows the mass-weighted equivalent,
\begin{equation}
    (\Delta Q)^2 = \sum_{\alpha i}M_{\alpha}\left(R^{\mathrm{exc}}_{\alpha i}-R^{\mathrm{GS}}_{\alpha i}\right)^2,
\end{equation}
where $R^{\mathrm{GS}}_{\alpha i}$ and $R^{\mathrm{exc}}_{\alpha i}$ denote the Cartesian coordinate $i \in \{x,y,z\}$ of atom $\alpha$ in 
the electronic ground state ($4f$) and the lowest excited state ($5d$), respectively, and $M_{\alpha}$ is the atomic mass of species $\alpha$.
Tables~\ref{tab:optical}--\ref{tab:second_shell} summarize the key structural and luminescent properties of all 14 configurations.

% ============================================================
% TABLE 1 — luminescent properties
% ============================================================
\begin{table}[t]
    \centering
    %\footnotesize
    \begin{tabular}{lcccccc}
    \toprule
    Label & $z$ & $\Delta E$ (eV) & $S^{\rm em}$ & $E^{\rm ZPL}$ (eV) & $E^{\rm em}$ (eV) \\
    \hline
    1O\_1             & 0.1875 & 0.000 & 2.196 & 2.676 & 2.635 \\
    1O\_2             & 0.1875 & 0.028 & 2.210 & 2.675 & 2.634 \\
    1O\_3             & 0.1875 & 0.040 & 2.082 & 2.699 & 2.657 \\
    1O\_4             & 0.1875 & 0.047 & 2.388 & 2.693 & 2.655 \\
    \hline
    2O\_1             & 0.25   & 0.000 & 2.293 & 2.670 & 2.630 \\
    2O\_2             & 0.25   & 0.027 & 2.155 & 2.690 & 2.649 \\
    2O\_3             & 0.25   & 0.346 & 2.200 & 2.700 & 2.657 \\
    2O\_4             & 0.25   & 0.563 & 2.288 & 2.661 & 2.620 \\
    \hline
    3O\_1             & 0.3125 & 0.000 & 2.277 & 2.697 & 2.655 \\
    3O\_2             & 0.3125 & 0.237 & 2.445 & 2.687 & 2.649 \\
    3O\_3             & 0.3125 & 0.246 & 2.354 & 2.655 & 2.615 \\
    3O\_4             & 0.3125 & 0.271 & 2.253 & 2.661 & 2.620 \\
    \hline
    1\_Al\_off\_plane & 0.1875 & 0.100 & 3.647 & 2.659 & 2.633 \\
    Al\_O\_far        & 0.1875 & 1.147 & 1.520 & 2.645 & 2.596 \\
    \hline
    \end{tabular}
    \caption{Luminescent properties of all studied configurations. $\Delta E$ is the energy relative to the lowest-energy structure at the same $z$. 
    $S^{\rm em}$, $E^{\rm ZPL}$, and $E^{\rm em}$ are computed from the 1D configuration coordinate model.}
    \label{tab:optical}
\end{table}

% ============================================================
% TABLE 2 — Eu–N distances + structural descriptors
% ============================================================
\begin{table*}[t]
    \centering
    %\scriptsize
    \setlength{\tabcolsep}{4pt}
    \begin{tabular}{lccccccccccc}
    \toprule
    Label & $d_1$ & $d_2$ & $d_3$ & $d_4$ & $d_5$ & $d_6$ & $d_7$ & $d_8$ & $d_9$ & $\langle d \rangle$ & $\sigma$ \\
    \hline
    1O\_1             & 2.486 & 2.495 & 2.523 & 2.748 & 2.748 & 2.788 & 2.788 & 2.829 & 2.829 & 2.693 & 0.138 \\
    1O\_2             & 2.469 & 2.510 & 2.524 & 2.771 & 2.771 & 2.777 & 2.777 & 2.818 & 2.818 & 2.693 & 0.138 \\
    1O\_3             & 2.476 & 2.520 & 2.522 & 2.768 & 2.768 & 2.785 & 2.785 & 2.816 & 2.816 & 2.695 & 0.135 \\
    1O\_4             & 2.492 & 2.497 & 2.531 & 2.744 & 2.744 & 2.797 & 2.797 & 2.827 & 2.827 & 2.695 & 0.137 \\
    \hline
    2O\_1             & 2.481 & 2.507 & 2.520 & 2.764 & 2.764 & 2.789 & 2.789 & 2.818 & 2.818 & 2.694 & 0.137 \\
    2O\_2             & 2.488 & 2.517 & 2.517 & 2.772 & 2.772 & 2.784 & 2.784 & 2.816 & 2.816 & 2.696 & 0.135 \\
    2O\_3             & 2.498 & 2.507 & 2.525 & 2.779 & 2.779 & 2.807 & 2.807 & 2.838 & 2.838 & 2.709 & 0.142 \\
    2O\_4             & 2.483 & 2.488 & 2.524 & 2.750 & 2.750 & 2.794 & 2.794 & 2.824 & 2.824 & 2.692 & 0.140 \\
    \hline
    3O\_1             & 2.503 & 2.511 & 2.521 & 2.797 & 2.797 & 2.807 & 2.807 & 2.824 & 2.824 & 2.710 & 0.141 \\
    3O\_2             & 2.488 & 2.515 & 2.524 & 2.754 & 2.754 & 2.805 & 2.805 & 2.814 & 2.814 & 2.697 & 0.135 \\
    3O\_3             & 2.478 & 2.500 & 2.521 & 2.767 & 2.767 & 2.794 & 2.794 & 2.814 & 2.814 & 2.694 & 0.139 \\
    3O\_4             & 2.483 & 2.504 & 2.514 & 2.769 & 2.769 & 2.782 & 2.782 & 2.821 & 2.821 & 2.694 & 0.138 \\
    \hline
    1\_Al\_off\_plane & 2.493 & 2.510 & 2.513 & 2.769 & 2.772 & 2.781 & 2.784 & 2.790 & 2.814 & 2.692 & 0.132 \\
    Al\_O\_far        & 2.481 & 2.489 & 2.496 & 2.758 & 2.758 & 2.764 & 2.764 & 2.765 & 2.765 & 2.671 & 0.129 \\
    \hline
    \end{tabular}
    \caption{Eu--N first-shell distances (sorted, in \AA) for all studied configurations, together with the mean $\langle d \rangle$ and standard deviation $\sigma$.}
    \label{tab:eu_n_distances}
\end{table*}

% ============================================================
% TABLE 3 — Second shell composition
% ============================================================
\begin{table}[t]
    \centering
    \begin{tabular}{lcccc}
    \toprule
    Label & Al & N & O & Si \\
    \hline
    1O\_1             & 2 & 25 & 1 & 13 \\
    1O\_2             & 2 & 25 & 1 & 13 \\
    1O\_3             & 2 & 23 & 1 & 13 \\
    1O\_4             & 2 & 25 & 1 & 13 \\
    \hline
    2O\_1             & 2 & 24 & 2 & 13 \\
    2O\_2             & 2 & 22 & 2 & 13 \\
    2O\_3             & 3 & 22 & 2 & 12 \\
    2O\_4             & 2 & 25 & 1 & 13 \\
    \hline
    3O\_1             & 3 & 21 & 3 & 12 \\
    3O\_2             & 2 & 24 & 2 & 13 \\
    3O\_3             & 2 & 22 & 2 & 13 \\
    3O\_4             & 2 & 22 & 2 & 13 \\
    \hline
    1\_Al\_off\_plane & 2 & 24 & 1 & 13 \\
    Al\_O\_far        & 0 & 26 & 0 & 15 \\
    \hline
    \end{tabular}
    \caption{Second-shell composition around Eu for all studied configurations, giving the number of Al, N, O, and Si neighbors in a radius of 5\AA from the Eu atom.}
    \label{tab:second_shell}
\end{table}

\begin{figure}[t]
    \centering
    \includegraphics[width=0.99\linewidth]{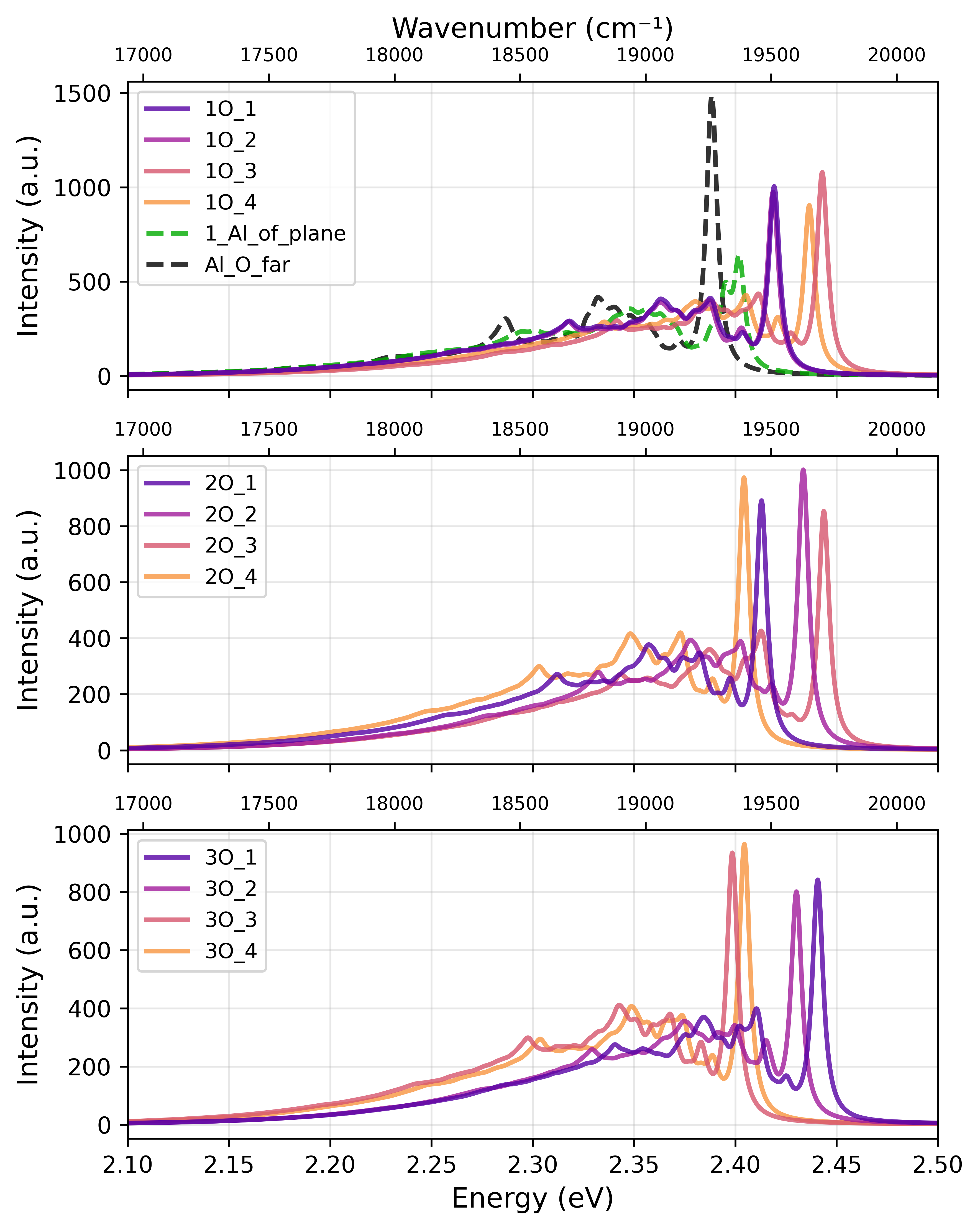} 
    \caption{Computed photoluminescence lineshapes at $T = 6$~K for all representative structural models.}
    \label{fig:SI_PL_all}
\end{figure}

\begin{figure}[t]
    \centering
    \includegraphics[width=0.99\linewidth]{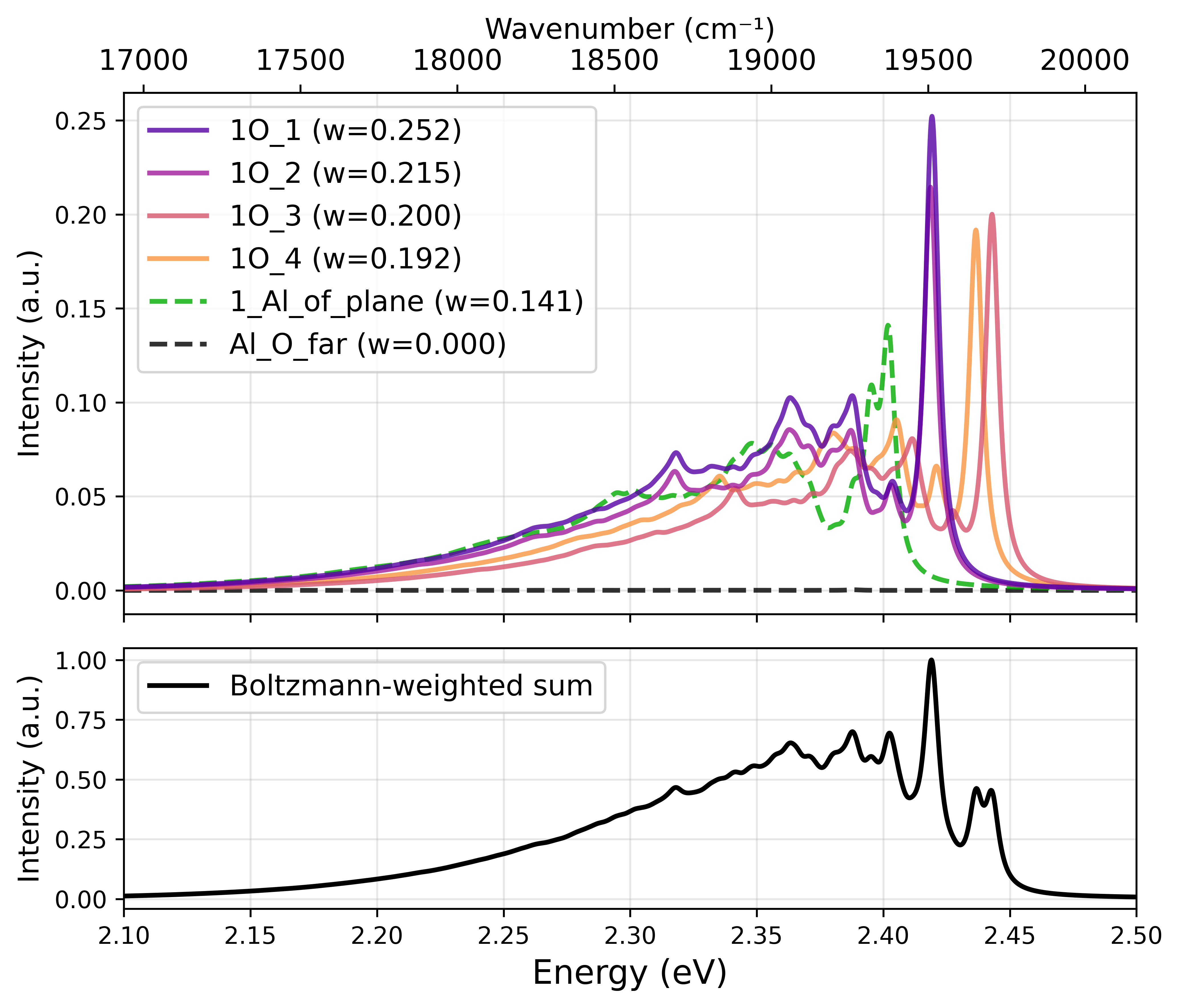}  
    \caption{Top: computed photoluminescence lineshapes at $T = 6$~K for the 1O case, weighted by their corresponding Boltzmann weights with a temperature of 2000K (synthesis-relevant temperature). 
    Bottom: Sum of the individual photoluminescence lineshapes, weighted by their Boltzmann weights. }
    \label{fig:SI_PL_all_weights}
\end{figure}

Table~\ref{tab:optical} reports the total energy relative to the lowest-energy configuration at the same composition ($\Delta E$), the emission Huang-Rhys
factor $S^{\rm em}$, the zero-phonon line energy $E^{\rm ZPL}$, and the estimated emission energy $E^{\rm em}$, all derived from the 1D configuration coordinate model~\cite{jia2017first, bouquiaux2021importance}. 
Across all 12 representative models, $E^{\rm ZPL}$ and $E^{\rm em}$ vary only weakly with composition and local arrangement
($\Delta E^{\rm ZPL} \lesssim 50$~meV), consistent with the robustness of the lineshape to the precise Al/O distribution observed in the main text.
The configuration that we name \texttt{1\_Al\_off\_plane} yields a larger value ($S^{\rm em} = 3.65$) due to the out-of-plane displacement of the Eu atom, while \texttt{Al\_O\_far} gives the smallest ($S^{\rm em} = 1.52$), reflecting the reduced electron-phonon coupling when the Al$_3$O cluster is far from Eu.

Table~\ref{tab:eu_n_distances} lists the nine Eu--N first-shell distances for each configuration, sorted in ascending order, together with their mean
$\langle d \rangle$ and standard deviation $\sigma$. 
The distances are divided into two groups: three short bonds ($d \approx 2.47$--$2.53$~\AA), associated to N atoms in the same [001] plane than Eu, 
and six longer ones ($d \approx 2.74$--$2.84$~\AA).
Table~\ref{tab:second_shell} gives the second-shell composition (number of Al, N, O, and Si atoms within $\sim$5~\AA\ of Eu).

Figure~\ref{fig:SI_PL_all} shows the computed PL lineshapes at $T = 6$~K for all 14 configurations, grouped by composition. 
Calculations are done in the 225-atoms supercell and with machine-learned \textsc{MatterSim} phonons. 
Within each composition group, the overall lineshape is well preserved across configurations. 
The ZPLs are aligned as in Fig.~\ref{fig:SI_Tdep}. 
The main variation between models at the same $z$ is a rigid shift of the ZPL, reflecting the small but non-negligible sensitivity of $E^{\rm ZPL}$ to the precise Al/O arrangement discussed in the main text. 
In particular, for the $z = 0.1875$ composition, the ZPL of 1O\_3 and 1O\_4 is blue-shifted by $\approx 25$~meV relative to 1O\_1 and 1O\_2, producing a
secondary ZPL peak at higher energy that corresponds to the shoulder experimentally identified as the ``other ZPL'' in
Fig.~5 of the main text. 
The two energetically-unfavoured structures (1\_Al\_off\_plane and Al\_O\_far) are shown in the top panel; their ZPL
positions and sideband shapes deviate more from the representative models.

Figure~\ref{fig:SI_PL_all_weights} further illustrates the $z = 0.1875$ case with Boltzmann-weighted lineshapes (with synthesis temperature $T = 2000$~K).
The top panel shows the individual contributions scaled by their respective weights. 
The bottom panel shows the resulting Boltzmann-weighted sum, in which the distinct ZPL peaks are clearly seen, supporting the assignment of the experimental shoulder to higher-energy structural variants.

\clearpage
% References
\medskip
\bibliography{main}